\documentclass[aps,prd,superscriptaddress,nofootinbib,amsmath,amsfonts]{revtex4-2}
\pdfoutput=1
\usepackage{setspace}
\usepackage{graphicx}
\usepackage{subfig}
\usepackage{epsfig}
\usepackage{bm}
\usepackage{amssymb}
\usepackage{float}  
\usepackage{amsmath}
\usepackage{dcolumn}
\usepackage[colorlinks]{hyperref}
\usepackage[usenames,dvipsnames]{color}
\usepackage{enumitem}

\hypersetup{ breaklinks=true, pdfstartview={FitH}, colorlinks=true, linkcolor=blue, citecolor=red, filecolor=magenta, urlcolor=blue, anchorcolor=green, linktocpage=true }

\newcommand{\nn}{\nonumber}

\def\doi{http://doi.org}

\newcommand{\be}{\begin{equation}}
\newcommand{\ee}{\end{equation}}
\newcommand{\beano}{\begin{eqnarray*}}
\newcommand{\eeano}{\end{eqnarray*}}
\newcommand{\ba}{\begin{eqnarray}}
\newcommand{\ea}{\end{eqnarray}}

\begin{document}

\title{Observational constraints and cosmographic analysis of  
$f({T},{T}_{{G}})$ gravity and cosmology}

\author{Harshna Balhara}
\email{harshna.ma19@nsut.ac.in}
\affiliation{Department of Mathematics, Netaji Subhas University of Technology, 
New Delhi-110 078, India}

\author{J. K. Singh}
\email{jksingh@nsut.ac.in}
\affiliation{Department of Mathematics, Netaji Subhas University of Technology, 
New Delhi-110 078, India}

\author{Shaily} 
\email{shailytyagi.iitkgp@gmail.com}
\affiliation{Department of Mathematics, Netaji Subhas University of Technology, New Delhi-110 078, India}
\affiliation{School of Computer Science Engineering and Technology, Bennett University, Greater Noida, India}

\author{Emmanuel N. Saridakis}
\email{msaridak@noa.gr}
\affiliation{National Observatory of Athens, Lofos Nymfon, 11852 Athens, Greece}
\affiliation{CAS Key Laboratory for Research in Galaxies and Cosmology,  
 University of Science and Technology of China, Hefei, Anhui 230026, China}
\affiliation{Departamento de Matem\'{a}ticas, Universidad Cat\'{o}lica del 
Norte, Avda. Angamos 0610, Casilla 1280 Antofagasta, Chile}

\begin{abstract}
 
We perform observational confrontation and cosmographic analysis of $f(T, T_G)$ 
gravity and cosmology. This  higher-order torsional gravity is based on both the 
torsion scalar, as well as on the teleparallel equivalent of the Gauss-Bonnet 
combination, and gives rise to an effective dark-energy 
sector which depends on the extra torsion contributions. We employ observational 
data from the Hubble function and Supernova Type Ia Pantheon datasets, applying 
a Markov Chain Monte Carlo sampling technique, and we provide the  
iso-likelihood contours, as well as the best-fit values for the parameters of 
the power-law model. Additionally, we reconstruct   the effective 
dark-energy equation-of-state parameter, which exhibits a quintessence-like 
behavior, while in the future the Universe enters into the phantom regime 
before it tends asymptotically to the cosmological constant value. Furthermore, 
we perform a detailed cosmographic analysis, examining the deceleration, jerk, 
snap, and lerk parameters, showing that the transition to acceleration occurs in 
the redshift range $ 0.53 < z_{tr} < 0.87 $, as well as the preference of 
the scenario for quintessence-like behavior. Finally, we apply the Om 
diagnostic analysis, as a cross-verification of the obtained behavior.  \\ 

 
\end{abstract}


\maketitle
 
\section{Introduction}
 
In order to describe the two phases of acceleration in the Universe's history, 
one can follow two main directions: early and late times. 
The first is to maintain general relativity as the underlying gravitational theory 
and introduce the dark-energy sector \cite{Copeland:2006wr,Cai:2009zp}  and/or 
the inflaton field   \cite{Bassett:2005xm}. The second is to construct 
gravitational modifications \cite{CANTATA:2021ktz,Capozziello:2011et}, which 
possess general relativity as a particular limit but in general provide the 
extra degrees of freedom that can lead to richer behavior. Note that the second 
direction can also alleviate the various cosmological tensions 
\cite{Abdalla:2022yfr} as well as bring the gravitational theory closer to a 
quantum description \cite{AlvesBatista:2023wqm}.
 
One of the first classes of modified gravity theories is obtained starting from the 
Einstein--Hilbert Lagrangian and adding new terms, yielding   $f(R)$ 
gravity \cite{Starobinsky:1980te, DeFelice:2010aj}, $f(G)$
gravity \cite{Nojiri:2005jg,DeFelice:2009aj}, $f(P)$ 
gravity \cite{Erices:2019mkd}, Lovelock gravity \cite{Lovelock:1971yv},    
   etc. Nevertheless, one could proceed 
beyond the standard, curvature-based formulation of gravity and use other 
geometrical quantities, such as torsion and non-metricity. In particular, one 
may start from the so-called teleparallel equivalent of general relativity 
\cite{Pereira.book,Maluf:2013gaa}, which uses the torsion scalar $T$ as a 
Lagrangian, and extend it to $f(T)$ gravity \cite{Bengochea:2008gz,Cai:2015emx}, 
 $f(R, T)$ gravity \cite{Singh:2024ckh, Shaily:2024nmy,Singh:2024kez, Singh:2018xjv}, $ f(R, L_m) $ gravity \cite{Singh:2024gtz}, $f(R,G)$ gravity \cite{Singh:2022gln,Singh:2022jue, Shaily:2024rjq}, etc.

 Alternatively, one can use the symmetric teleparallel theory, which uses the non-metricity scalar as the 
Lagrangian \cite{Nester:1998mp}, and extend it to $f(Q)$ gravity 
\cite{BeltranJimenez:2017tkd,Heisenberg:2023lru, Goswami:2023knh}. All these classes of 
gravitational modification has been shown to lead to very rich cosmological 
behavior, and thus, have attracted the interest of the community 
\cite{Cognola:2006eg, Nojiri:2010wj,delaCruz-Dombriz:2011oii,
Harko:2011kv,Bamba:2012cp, Clifton:2011jh, Capozziello:2013vna,Skugoreva:2014ena,Gleyzes:2014qga,Arai:2017hxj, Hohmann:2018wxu,BeltranJimenez:2018vdo,Lu:2019hra,Singh:2019fpr,
 BeltranJimenez:2019tme,Lazkoz:2019sjl, Yan:2019gbw,
Ayuso:2020dcu,Barros:2020bgg,Kovacs:2020ywu,Caruana:2020szx,Hohmann:2021ast, 
DAmbrosio:2021zpm, Anagnostopoulos:2021ydo,Atayde:2021pgb,Lin:2021uqa,Khyllep:2021pcu, 
Esposito:2021ect, DAmbrosio:2021pnd,Moreira:2021xfe, Zhao:2021zab, 
De:2022wmj,Chatzifotis:2022mob,Shabani:2023xfn, Singh:2022ptu, 
Capozziello:2022wgl,Chatzifotis:2023ioc,Koussour:2023gip,Karakasis:2023hni,
Basilakos:2023xof, Bakopoulos:2023tso,Boehmer:2023knj, Singh:2023gxd}. Finally, 
it is interesting to mention that one can also add boundary  terms in the above formulations, resulting in 
 $f(T,B)$ gravity \cite{Bahamonde:2015zma}, and in  $ f(Q,C) $ gravity 
\cite{De:2023xua}.
 
Since in curvature gravity one can use higher-order invariants  in the 
Lagrangian, an interesting question is whether one can use such invariants in 
torsional gravity too. Indeed, as was shown in \cite{Kofinas:2014owa}, within the teleparallel formulation of gravity it is also possible to incorporate higher-order  corrections. In 
particular, one can construct the teleparallel equivalent of the Gauss--Bonnet 
term, namely, ${T}_{{G}}$, and then, use it to formulate the general class of 
$f({T}, {T}_{{G}})$ gravity \cite{Kofinas:2014owa,Kofinas:2014aka}, which is also known to 
have very interesting cosmological applications  
\cite{Kofinas:2014daa,Azhar:2020coz,Capozziello:2016eaz,
delaCruz-Dombriz:2017lvj,delaCruz-Dombriz:2018nvt,Chattopadhyay:2014xaa,
Zubair:2015yma,Sharif:2017pvo, Mustafa:2019eet, 
Asimakis:2021yct,Lohakare:2022hvf, 
Bahamonde:2016kba, Kadam:2022daz}. 
 
In all classes of modified gravity that include an unknown function $f$, the 
main task is to exactly determine the form of this function, as well as to 
constrain the range of the involved parameters. In order to achieve this, one 
may use theoretical considerations, such as impose the validity of various 
symmetries \cite{Capozziello:2007wc,Paliathanasis:2014iva}; however, the most 
powerful tool is to leave the involved function free and use observational data 
in order to extract observational constraints. Hence,  in this work we are 
interested in performing such observational analysis in the case of  $f({T}, 
{T}_{{G}})$ gravity and cosmology. In particular, we use data from Hubble 
constant measurements from  cosmic chronometers (CCs), from supernova Type Ia 
(SNIa Pantheon dataset) observations, as well as from baryon acoustic 
oscillation  (BAO)  observations. Additionally, we study the evolution of 
various cosmographic parameters, and we perform the Om diagnostic.
 
The plan of the work is the following. In Section~\ref{section-II},  we review 
 $f({T}, {T}_{{G}})$ gravity and we apply it in a cosmological framework. In 
Section~\ref{Observationalconstraints}, we present the observational datasets 
and we perform the observational confrontation, providing the corresponding 
contour plots. In Section \ref{section-IV}, we perform a cosmographic analysis 
and we apply the $Om(z)$ diagnostic. Finally, we summarize our findings and 
conclusions in Section \ref{section-V}.
\section{\boldmath{$f({T},{T}_{{G}})$} Gravity and Cosmology}
\label{section-II}

In this section, we   briefly review  $f(T,T_G)$ gravity, and then, we apply it in 
a cosmological framework.

\subsection{$f({T},{T}_{{G}})$ Gravity}

In the torsional formulation of gravity one uses the  vierbein field $e_a(x^\mu)$ 
as the dynamical variable, expressed in terms of coordinate components as
$e_a=e^{\,\,\, \mu}_a\partial_\mu$ (Greek indices   run over the coordinate
spacetime and Latin indices run over  the tangent space). Additionally,  
one uses the
Weitzenb{\"{o}}ck  
connection   1-form,  that defines the parallel transportation,   which in 
all coordinate frames is written as
$\omega_{\,\,\,\mu\nu}^{\lambda}=e_{a}^{\,\,\,\lambda}e^{a}_{\,\,\,\mu
, \nu } $.  Moreover, we mention that its  tangent-space
components are $\omega_{\,\,\,bc}^{a}=0$, assuring the property of zero 
non-metricity. For an orthonormal vierbein the metric   is expressed as
\begin{equation}
\label{metrdef}
g_{\mu\nu} =\eta_{ab}\, e^a_{\,\,\,\mu}  \, e^b_{\,\,\,\nu},
\end{equation}
with $\eta_{ab}=\text{diag}(-1,1,1,1)$ ($a,b,...$   indices are
raised/lowered using $\eta_{ab}$).

The torsion  tensor is defined as \cite{Pereira.book,Maluf:2013gaa}
\begin{equation}
T^{\lambda}_{\,\,\,\mu\nu}=
e_{a}^{\,\,\,\lambda}\left(\partial_\nu e^{a}_{\,\,\,\mu}-\partial_\mu
e^{a}_{\,\,\,\nu}\right),
\end{equation} 
while  the contorsion tensor, which equals the difference
between the Weitzenb%
\"{o}ck and Levi--Civita connections, is  
$
\mathcal{K}^{\mu\nu}_{\:\:\:\:\rho}=-\frac{1}{2}\Big(T^{\mu\nu}_{
\:\:\:\:\rho}
-T^{\nu\mu}_{\:\:\:\:\:\rho}-T_{\rho}^{\:\:\mu\nu}\Big)$. Hence, one can define 
the torsion scalar as 
\begin{eqnarray}
T&=&\frac{1}{4}T^{\mu\nu\lambda}T_{\mu\nu\lambda}+\frac{1}{2}T^{\mu\nu\lambda}
T_{\lambda\nu\mu}-T_{\nu}^{\,\,\,\nu\mu}T^{\lambda}_{\,\,\,\lambda\mu},
\label{Tquad}
\end{eqnarray}
which   is then used to construct the action of the theory as 
\begin{eqnarray}
S=-\frac{1}{2\kappa^2}\int d^4 x \,e\,T \, ,
\label{teleaction}
\end{eqnarray}
with $e=\det{(e^{a}_{\,\,\,\mu})}=\sqrt{|g|}$ and $\kappa^2\equiv 8\pi
G$  the   gravitational constant. The above theory  is called the teleparallel
equivalent of general relativity, since variation in terms of the vierbein 
gives rise to exactly the same equations as general relativity. Finally, 
upgrading $T$ to an arbitrary function $f(T)$, namely, writing the action 
\cite{Cai:2015emx}
\begin{eqnarray}
S=\frac{1}{2\kappa^2}\int d^4 x \,e\,f(T) \, ,
\label{teleaction}
\end{eqnarray}
gives rise to the simplest torsional modified gravity, i.e., $f(T)$ gravity.

Similarly to the fact that one can use the Riemann tensor in
order to build higher-order curvature invariants such as the 
Gauss--Bonnet term  $
G=R^{2}-4R_{\mu\nu}R^{\mu\nu}+R_{\mu\nu\kappa\lambda}R^{\mu\nu\kappa\lambda}
$, one can   use the torsion tensor in order to build higher-order
torsion invariants. Specifically, one can construct 
\cite{Kofinas:2014owa}  
  \begin{eqnarray}
&&
\!\!\!\!\!\!\!\!\!\!\!\!\!
T_G=\left(\mathcal{K}^{\kappa}_{\,\,\,\varphi\pi}\mathcal{K}^{\varphi\lambda}_
{\,\,\,\,\,
\,\, \rho }\mathcal{K}^{\mu}_{\,\,\,\,\chi\sigma}
\mathcal{K}^{\chi\nu}_{\,\,\,\,\,\,\,\tau}
-2\mathcal{K}^{\kappa\!\lambda}_{\,\,\,\,\,\,\pi}\mathcal{K}^{\mu}_{
\,\,\,\varphi\rho}
\mathcal{K}^{\varphi}_{\,\,\,\chi\sigma}\mathcal{K}^{\chi\nu}_{\,\,\,\,\,\,\tau}
\right.
\nn\\
&&\left.  \!\!\!\!\!\!\!\!\!\!\!\!\!
+2\mathcal{K}^{\kappa\!\lambda}_{\,\,\,\,\,\,\pi}\mathcal{K}^{\mu}_{
\,\,\,\,\varphi\rho}
\mathcal{K}^{\varphi\nu}_{\,\,\,\,\,\,\chi}\mathcal{K}^{\chi}_{\,\,\,\,
\sigma\tau}
+2\mathcal{K}^{\kappa\!\lambda}_{\,\,\,\,\,\,\pi}\mathcal{K}^{\mu}_{
\,\,\,\,\varphi\rho}
\mathcal{K}^{\varphi\nu}_{\,\,\,\,\,\,\,\sigma,\tau}\right)
\delta^{\pi\rho\sigma\tau}_{\kappa \lambda \mu \nu}
\label{TG},
\end{eqnarray}
where  the generalized
$\delta^{\pi\rho\sigma\tau}_{\kappa \lambda \mu \nu}$ is the determinant
of the Kronecker deltas.
The above term  is   the teleparallel equivalent
of the Gauss--Bonnet combination, since $T_G$ and $G$ differ only by a boundary 
term. Thus, although using $T_G$ as a Lagrangian will give the same field 
equations as using $G$ in curvature gravity (zero in four dimensions since 
both terms are topological invariants), $f(T_G)$ will lead to  different 
equations than $f(G)$. In summary, one can construct the general theory of 
$f(T,T_G)$ gravity, writing the action
  \cite{Kofinas:2014owa}
\begin{eqnarray}
S =\frac{1}{2\kappa^{2}}\!\int d^{4}x\,e\,f(T,T_G)\,.
\label{fGBtelaction}
\end{eqnarray}
{Such} 
 a theory  is  different from $f(R,G)$ gravity  
\cite{Nojiri:2005jg,DeFelice:2008wz,Jawad:2013uil}, as expected. 
Finally, let us comment here that the various modified theories of 
gravity can be re-expressed as usual metric theories plus additional 
degrees of freedom. The corresponding Hamiltonian  analysis of the number of 
degrees of freedom of classes of torsional gravities,   as well as their 
corresponding physics, have been 
extensively studied in the literature (for instance, see the reviews 
\cite{Cai:2015emx,Krssak:2018ywd,Bahamonde:2021gfp}). Nevertheless, the 
important point is 
that the interpretation, and thus, the justification in each theory is different.

\subsection{$f(T,T_G)$ Cosmology}
\label{fTcosmology}

Let us now apply $f(T, T_G)$ gravity in a cosmological framework. We consider  a 
spatially flat Friedmann--Robertson--Walker (FRW) metric of the form
\begin{equation}
ds^{2}=-dt^{2}+a^{2}(t)\delta_{\hat{i}\hat{j}}dx^{\hat{i}}dx^{\hat{j}}\,,
\label{metriccosmo}
\end{equation}
with $a(t)$   the scale factor, which can arise from the diagonal vierbein
\begin{equation}
\label{vierbeincosmo}
e^{a}_{\,\,\,\mu}=\text{diag}(1,a(t),a(t),a(t))
\end{equation}
through (\ref{metrdef}). Therefore, inserting  (\ref{vierbeincosmo}) into 
  (\ref{Tquad}) and  (\ref{TG})  we acquire 
\begin{eqnarray}
\label{Tcosmo1}
 &&T=6H^2\\
 &&T_G= 24H^2\big(\dot{H}+H^2\big) ,
 \label{TGcosmo1}
\end{eqnarray}
where $H=\frac{\dot{a}}{a}$ is the Hubble function and with dots denoting
differentiation with respect to $t$.
Lastly,  we add the standard matter $S_m$, corresponding to   a perfect 
fluid   of energy
density $\rho_m$ and pressure $p_m$.

Variation in the total action $S+S_m$ leads to the following   
  Friedmann equations \cite{Kofinas:2014owa}:
\begin{equation}
f-12H^{2}f_{T}-T_G f_{T_G}
+24H^{3}\dot{f_{T_G}}=2\kappa^{2}\rho_m
\label{Fr1}
\end{equation}
\begin{eqnarray}
&&
\! 
f-4\big(3H^2+\dot{H}\big)f_T-4H\dot{f_T}-T_G 
f_{T_G}\nonumber\\
&&
\ \  
+\frac{2}{3H}T_G\dot{f_{T_G}}+8H^2\ddot{f_{T_G}}
=-2\kappa^{2} p_m\,,
\label{Fr2}
\end{eqnarray}
where $\dot{f_{T}}=f_{TT}\dot{T}+f_{TT_{G}}\dot{T}_{G}$,
$\dot{f_{T_{G}}}=f_{TT_{G}}\dot{T}+f_{T_{G}T_{G}}\dot{T}_{G}$,
and $\ddot{f_{T_{G}}}=f_{TTT_{G}}\dot{T}^{2}+2f_{TT_{G}T_{G}}\dot{T}
\dot{T}_{G}+f_{T_{G}T_{G}T_{G}}\dot{T}_{G}^{\,\,2}+
f_{TT_{G}}\ddot{T}+f_{T_{G}T_{G}}\ddot{T}_{G}$,
with $f_{TT}$, $f_{TT_{G}}$, ... denoting partial differentiations
of $f$ with respect to $T$, $T_{G}$. Moreover, note that the  various 
time derivatives of
$\dot{T}$, $\ddot{T}$, $\dot{T}_{G}$, and $\ddot{T}_{G}$ are  
obtained using (\ref{Tcosmo1}) and (\ref{TGcosmo1}). Furthermore, 
 one can rewrite the Friedmann \mbox{Equations (\ref{Fr1}) and
(\ref{Fr2})} as
 \begin{eqnarray}
\label{Fr1b}
H^2& =& \frac{\kappa^2}{3}\left(\rho_m + \rho_{DE} \right)   \\
\label{Fr2b}
\dot{H}& =&-\frac{\kappa^2}{2}\left(\rho_m +p_m+\rho_{DE}+p_{DE}\right),
\end{eqnarray}
having introduced an  effective dark energy sector 
with energy density and pressure of the forms
\begin{equation}
\label{rhode}
\rho_{DE}\equiv\frac{1}{2\kappa^2}\left( 6H^2
-f+12H^{2}f_{T}+T_G f_{T_G}-24H^{3}\dot{f_{T_G}}\right)
\end{equation}
\begin{eqnarray}
\label{pde}
&&
\!\!\!\!\!\!\!\!\!\!\!\!\!\!\!\!\! p_{DE}  \equiv   \frac{1}{2\kappa^2}\Big[ 
-2(2\dot{H} + 3H^2)+f - 4\big(\dot{H} + 3H^2\big)f_T\nonumber\\
&&\ \ -4H\dot{f_T}-T_G
f_{T_G}+\frac{2}{3H}T_G\dot{f_{T_G}}+8H^2\ddot{f_{T_G}}
\Big],
\end{eqnarray}
and thus, the dark-energy equation-of-state parameter is defined as
\begin{equation}
\label{wDEdef}
 w_{DE} \equiv 
\frac{p_{DE}}{\rho_{DE}}.
\end{equation}
{Finally,}  note that the equations close by considering the matter conservation 
equation,
\begin{equation}
\label{cpnserv1}
 \dot{\rho}_m+3H(\rho_m+p_m)=0,
\end{equation}
which, inserting into the first Friedmann Equation (\ref{Fr1b}) and substituting 
into the second  one (\ref{Fr2b}), leads to the dark energy conservation, namely,
\begin{equation}
\label{cpnserv2}
 \dot{\rho}_{DE}+3H(\rho_{DE}+p_{DE})=0.
\end{equation}
{Lastly,}  as usual,  we introduce the density parameters  
  \begin{eqnarray}
 &&  \Omega_{m}\equiv \frac{\kappa^2\rho_m}{3  H^2}\\
   && \Omega_{DE}\equiv \frac{\kappa^2\rho_{DE}}{3 H^2}.
\end{eqnarray}

{The above equations determine the Universe's evolution in the framework of 
$f(T, T_G)$ cosmology. In this work, we are interested in confronting them with 
observational data and extracting constraints for the involved parameters. This 
is performed in the following sections.}


\section{Observational Constraints }
\label{Observationalconstraints}

We are interested in extracting observational constraints in the scenario of 
$f(T,T_G)$ gravity and cosmology. Firstly, we present the datasets that we 
employ in our investigation, as well as the corresponding methodology, and then, 
we 
perform the analysis for a specific $f(T,T_G)$ model.

\subsection{Datasets and Analysis }
\label{datasetsanalysis}
 
We explore the parameter space using a  Markov chain Monte Carlo (MCMC)  
sampling technique and predominantly rely on the emcee library in Python 
\cite{Foreman-Mackey:2012any} ({one} 
 could equally well use standard Monte 
Carlo or other 
Monte Carlo variants). In the following, we employ the newly 
published Pantheon dataset, comprising 1048 observations of supernova Type Ia 
(SNeIa) events gathered from various 
surveys, including Low-z, SDSS, SNLS, Pan-STARRS1 (PS1) Medium Deep Survey, and 
HST \cite{Pan-STARRS1:2017jku}. The dataset covers a redshift range of $z \in 
(0.01, 2.26)$. To focus on the evidence about the expansion history of 
the Universe, particularly the connection between distance and redshift, two 
distinct observational datasets are utilized to constrain the model 
being examined. Notably, recent research exploring the significance of $H(z)$ 
and SNeIa (Type Ia supernovae) data in cosmological constraints has revealed 
their ability to limit cosmic parameters.

\subsubsection{$Pantheon$ SNeIa Dataset}

The Pantheon probe includes a sample of 1048 Type Ia supernovae (SNeIa), and 
the $\chi^2_{Pan}$ function is defined as \cite{Pan-STARRS1:2017jku} 
\begin{equation}\label{12}
 \chi_{Pan}^{2}=\sum\limits_{i=1}^{1048}\left[ 
\frac{\mu_{th}(\mu_0,z_i)-\mu_{obs}(z_{i})}{\sigma_i}\right]^2.
\end{equation}
{Furthermore,} the symbol $\sigma_i$ denotes the standard error associated with 
the actual value of $H$. The theoretical distance modulus $\mu_{th}$ is defined 
as $\mu_{th}^i = \mu(D_l)=m-M= 5log_{10}D_l(z) + \mu_0$, where the  apparent 
magnitude is represented by $m$, the absolute magnitude by $M$, and the 
nuisance parameter $\mu_0$ is defined as $\mu_0 = 5log\frac{H_0^{-1}}{Mpc}+25$. 
Additionally, the luminosity distance $D_l(z)$ is defined as $D_l(z) = (1 + 
z)H_0\int\frac{1}{H(z')}dz'$. Finally, to approximate the limited 
series, the $H(z)$ series is truncated at the tenth term and integrated to 
obtain the luminosity 
distance.

\subsubsection{{$Hubble$} 
 Dataset}

In order to determine the expansion rate of the Universe at a specific redshift 
$z$, we employ the commonly used differential age (DA) method, the baryon 
acoustic oscillation  (BAO)  method, and 
other methods in the redshift range {$ 0.07 \leq z \leq 2.42$} 
 \cite{Sharov:2018yvz,Chimento:2007da}. This approach allows us to estimate 
$H(z)$ by utilizing the equation $(1 + z)H(z) = -\frac{dz}{dt}$. The average 
values of the model parameters, as well as of the present value of the matter 
density parameter  $\Omega_{m0}$, are obtained by 
minimizing the chi-square value. The chi-square function  based on Hubble data 
is expressed as  
\begin{equation}\label{11}  
\chi^2_{H}=\sum_i^{55}\frac{[H_{th}(z_i)-H_{obs}(z_i)]^2}{\sigma_i^2},
\end{equation}
where the standard error associated with the experimental values of the Hubble 
function is represented by $\sigma_i$. The terms $H_{th}(z_i)$ and 
$H_{obs}(z_i)$ correspond to the theoretical and observed values of the Hubble 
parameter, respectively.

 \subsubsection{$Hubble+Pantheon$ Dataset}
 
We employ the total likelihood function to obtain combined constraints for the 
model parameters by utilizing data from both the Hubble 
and Pantheon samples. As a result, the relevant chi-square function for this 
analysis is given by 
 \begin{equation}\label{13}
\chi_{joint}^2=\chi_H^2+\chi_{Pan}^2.
 \end{equation}

\subsection{Results} \label{ResultsObserv}

We now have all the material needed to perform the observational 
confrontation of   $f(T, T_G)$ gravity and cosmology. To move forward, we would need to have a specific ansatz.
As is usual in many modified gravity models, the aim is 
to find a phenomenological form that could replace the cosmological constant. 
Hence, in our work, we do not desire to consider an explicit cosmological 
constant, but choose an $f(T, T_G)$ form that could lead to viable cosmological 
behavior. In particular, in torsional theories the 
Lagrangian $-T+const.$ corresponds to the teleparallel equivalent of the general 
relativity, namely, it leads to the same equations as general 
relativity plus a cosmological constant, in the following we consider the 
$T$ term but instead of a constant we desire to examine whether the addition 
of a general term of the form   $ \alpha T_G ^\beta T 
^\eta$,  
where $\alpha$, $\beta$, and $\eta$ are   constants, can lead to viable 
cosmological phenomenology (such terms have been  
studied in detail 
in the literature   
\cite{Kofinas:2014daa,Azhar:2020coz,Capozziello:2016eaz,
delaCruz-Dombriz:2017lvj,delaCruz-Dombriz:2018nvt,Chattopadhyay:2014xaa,
Zubair:2015yma,Sharif:2017pvo, Mustafa:2019eet, 
Asimakis:2021yct,Lohakare:2022hvf, 
Bahamonde:2016kba, Kadam:2022daz}). Note that the coupling 
parameter $\alpha$  determines the scale in which torsional 
corrections become important   
\cite{Cai:2015emx,Krssak:2018ywd,Bahamonde:2021gfp}.

 Hence, in the following we
consider 
\begin{equation}
  f({T},{T}_{{G}}) = 
-{T}+\alpha {T}_{{G}}^\beta {T}^{\eta}. 
\end{equation}
 {In} this case, the 
Friedmann  Equations (\ref{Fr1b}) and  (\ref{Fr2b}), respectively, become 
\begin{eqnarray}\label{14}
 &&
 \!\!\!\!\!\! \!\!
 2 \kappa^2  \rho_m\! = \!   \alpha {T}^{\eta} 
{T}_{{G}}^\beta-{T}-12 H^2(\alpha \eta {T}^{\eta-1} 
{T}_{{G}}^\beta\!-\!1)-\alpha \beta {T}^{\eta} {T}_{ 
{G}}^{\beta}
\nonumber\\
&&\!\!\!\!\!\!
+24H^3\left[\eta \alpha\beta 
{T}^{\eta-1}{T}_{{G}}^{\beta-1}\dot{{T}}+\alpha 
\beta(\beta-1){T}^{\eta}{T}_{{G}}^{\beta-2}\dot{{T}_{{G}}}\right],
\end{eqnarray}
\begin{eqnarray}\label{15}
   &&\!\!\!\!\!\!\! 
   2\kappa^2    p_m= (12H^2+4\dot{H})(\eta 
\alpha{T}^{\eta-1}{T}_{{G}}^{\beta}-1)+{T}-\alpha {T}^{\eta} 
{T}_{{G}}^{\beta}
\nonumber\\
&& 
+8\alpha 
H({T}_{{G}}^{\beta}\dot{{T}}+\beta {T} 
\dot{{T}_{{G}}}{T}_{{G}}^{\beta-1})
+\alpha \beta {T}^{\eta} 
{T}_{{G}}^{\beta}\nonumber
\\ &&  
 -\frac{2}{3H}{T}_{{G}}\left[\eta\alpha\beta 
{T}^{\eta-1}{T}_{{G}}^{\beta-1}\dot{{T}}
   +\alpha 
\beta(\beta-1){T}^{\eta-1}{T}_{{G}}^{\beta-2}\dot{{T}_{{G}}}
\right]\nonumber
\\
&&  
-8H^2\Big[2\alpha \beta {T}_{{G}}^{\beta-1}\dot{{T}}^2+2\eta\alpha 
\beta(\beta-1){T}^{\eta-1}{T}_{{G}}^{\beta-2}\dot{{T}} 
\dot{{T}_{{G}}}\nonumber\\
&& \ \ \ \ \ \ \ \ +\alpha 
\beta(\beta-1)(\beta-2){T}^{\eta}{T}_{{G}}^{\beta-2}\dot{{T}}^2
  \nonumber
\\ 
&& \ \ \ \ \ \ \ \ +\eta \alpha \beta 
{T}^{\eta-1}\ddot{{T}}{T}_{{G}}^{\beta-1}+\alpha\beta(\beta-1){T}^{\eta}
\Big]{T}_{{G}}^{\beta-2}\ddot{{T}_{{G}}}  .
\end{eqnarray}

For convenience relating to observational confrontation, we use the redshift 
$z=-1+\frac{a_0}{a}$ as the independent variable, setting additionally the 
current scale factor to $a_0=1$.  Thus, to calculate the expansion 
rate $(1+z)H(z) = -\frac{dz}{dt}$, we express ${T}$ and ${T}_{{G}}$ as 
functions 
of the redshift parameter $z$ as 
\begin{eqnarray}\label{9}
   && \!\!\!\!\!\! \!\!\!\! {T}=6H_{0}^2E(z),\nonumber\\
&& \!\!\!\!\!\!\!\!\! \!{ 
T}_{{G}}=24H_{0}^2E(z)\Big[-\frac{H_0^2(1+z)E'(z)}{2}+H_0^2E(z)\Big],
\end{eqnarray}
where $E^2(z)=H^2(z)/H_0^2$ is the normalized squared Hubble function, with 
$H_0$ the Hubble parameter at present, and primes indicate the 
derivative with respect to the redshift.  
 
\begin{figure}[!htbp]
	   {\includegraphics[scale=0.53]{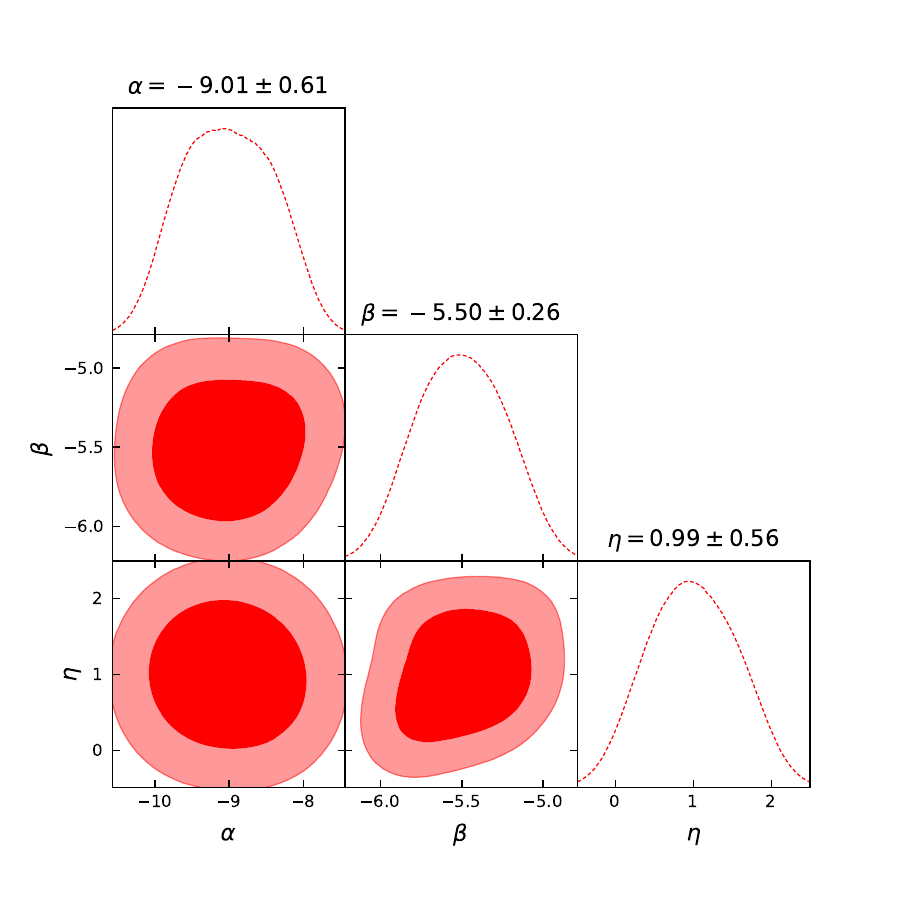}} \hfill
	   {\hspace{-1.5cm}\includegraphics[scale=0.28]{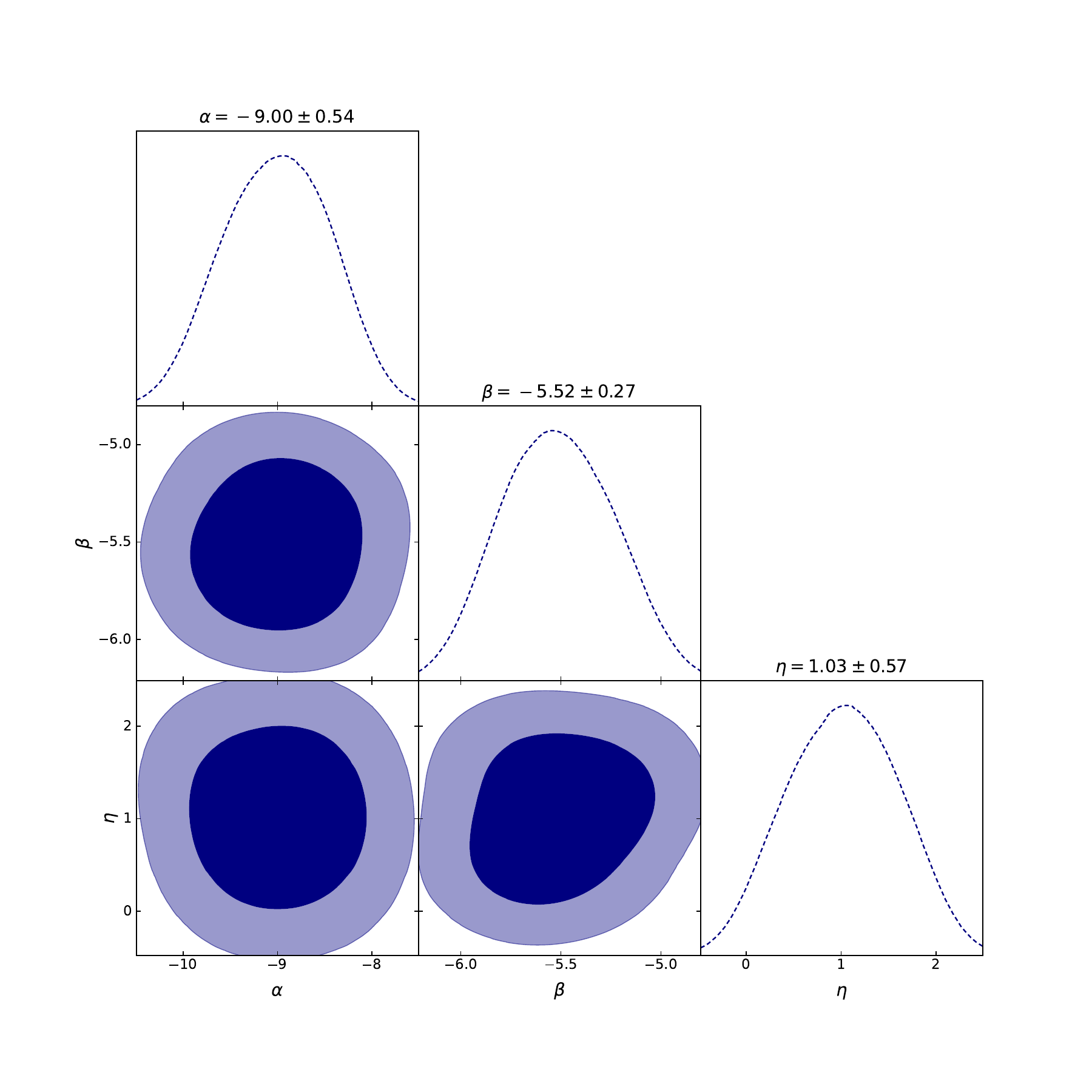}} \par\vspace{-12pt}
	   {\includegraphics[scale=0.53]{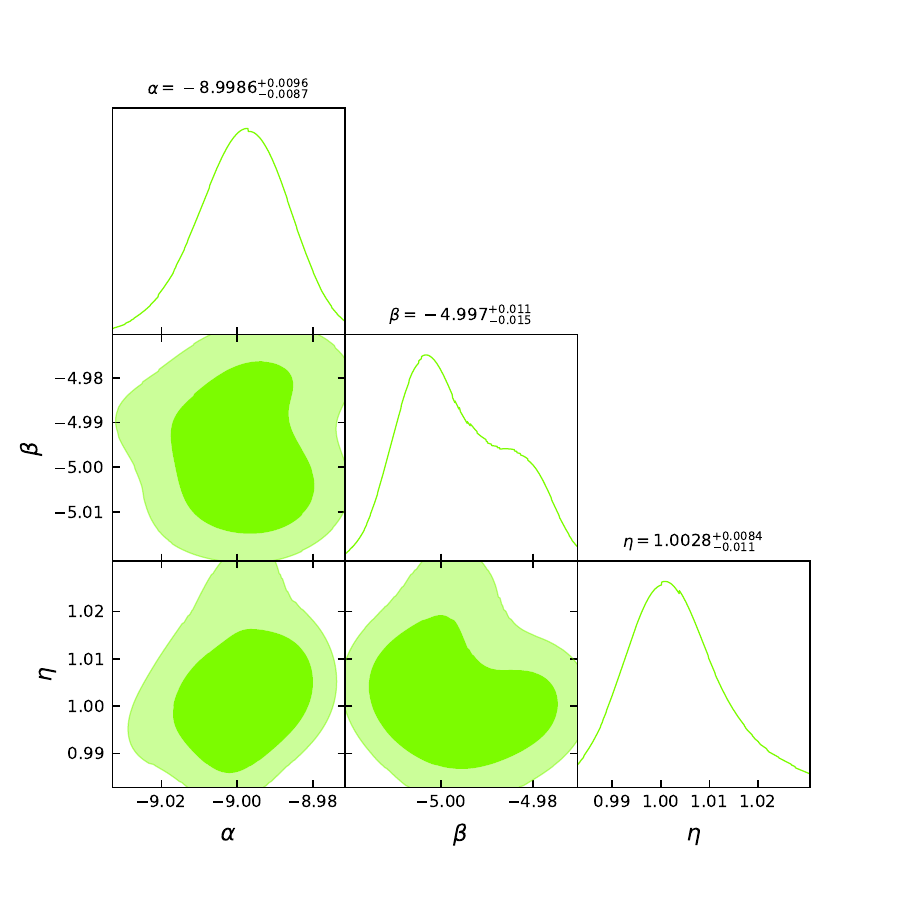}}
\caption{The $1\sigma$ and $2\sigma$  iso-likelihood contours for power-law $f(T,T_G)$  
cosmology, for the various  two-dimensional subsets 
of the parameter space, for various datasets. The upper graph corresponds 
to the $Hubble$ dataset, the middle  corresponds 
to the $Pantheon$ dataset, and the lower graph corresponds to the joint analysis
$Hubble+Pantheon$.}
 \label{Hz-Pan-Joint}
\end{figure}

We perform the observational analysis described in the previous sections,   
and in Table \ref{tabparm1} we provide the obtained results.
Additionally, in Figure~\ref{Hz-Pan-Joint}  we depict the 
$1-\sigma$ and $2-\sigma $ confidence regions in various two-dimensional 
projections, obtained through contour analyses of $\chi^2$ in 
the parameter space \cite{Singh:2022nfm, Singh:2023ryd}. 
We mention here that the obtained parameter er values of $\beta$ and 
$\eta$   can indeed lead to viable 
cosmological behavior  in agreement with the data, although an explicit 
cosmological constant is absent. Thus, we have indeed found a non-trivial 
$f(T,T_G)$  model that can lead to viable phenomenology.

Table \ref{tabparm1} contains the constrained values of the model parameters obtained from various datasets.  
\begin{table*} 
\caption{Observational constraints on $f(T,T_G)$ cosmology  using $Hubble$, $Pantheon$, and the joint 
analysis $Hubble+Pantheon$ datasets.}
\begin{center}
\begin{tabular}{l c  c c c r} 
\hline\hline
{Dataset} &  ~~~~~ $\Omega_{m0}$ &     ~~~~~ $\alpha $  &  ~~~~~ $ \beta $  & 
~~~~~  $ \eta $   
\\
\hline 
\\
{$ H(z) $ }     &  ~~~~~ $ 0.233^{+0.018}_{-0.015} $     &  ~~~~~ $ -9.01^{+0.61}_{-0.61} $   &  ~~~~~ $ -5.50^{+0.26}_{-0.26} $  &  ~~~~~ $ 0.99^{+0.56}_{-0.56} $       
\\
\\
{$ Pantheon $ }     &  ~~~~~ $ 0.297^{+0.064}_{-0.064} $   &  ~~~~~ $ -9.00^{+0.54}_{-0.54} $ & ~~~~~ $ 
-5.52^{+0.27}_{-0.27} $  &  ~~~~~ $ 1.03^{+0.57}_{-0.57} $   
\\
\\
{ H(z)+Pantheon }  &  ~~~~~$ 0.257^{+0.011}_{-0.011} $   &  ~~~~~$ -8.9986^{+0.0096}_{-0.0087} $  & ~~~~~ $ 
-4.997^{+0.011}_{-0.015} $  &  ~~~~~ $ 1.0028^{+0.0084}_{-0.011} $  
\\
\\
\hline\hline  
\end{tabular}    
\end{center}
\label{tabparm1}
\end{table*}

\begin{figure}[!htbp]
	{\includegraphics[scale=0.4]{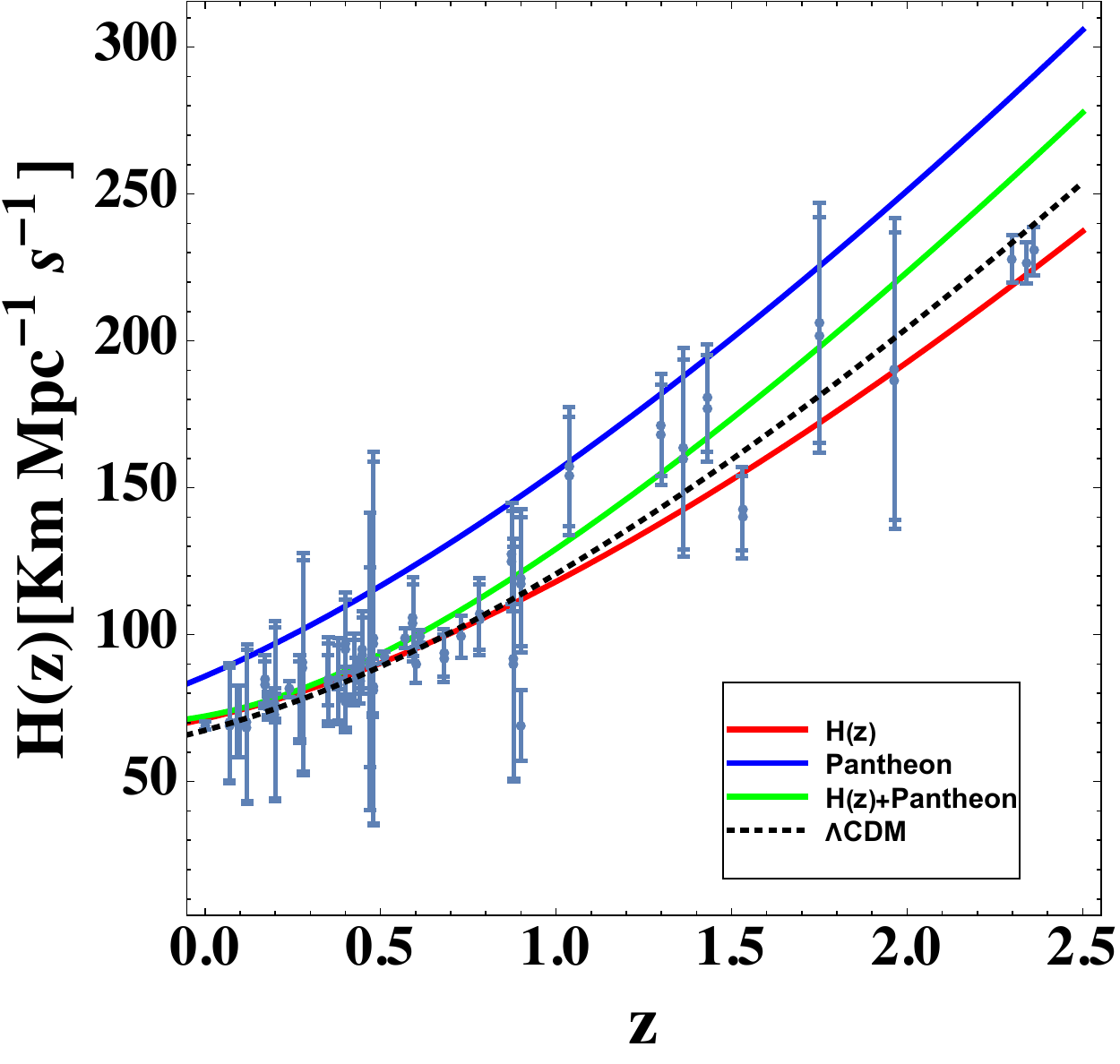}}\hfill
	   {\hspace{-1.5cm}
	\includegraphics[scale=0.38]{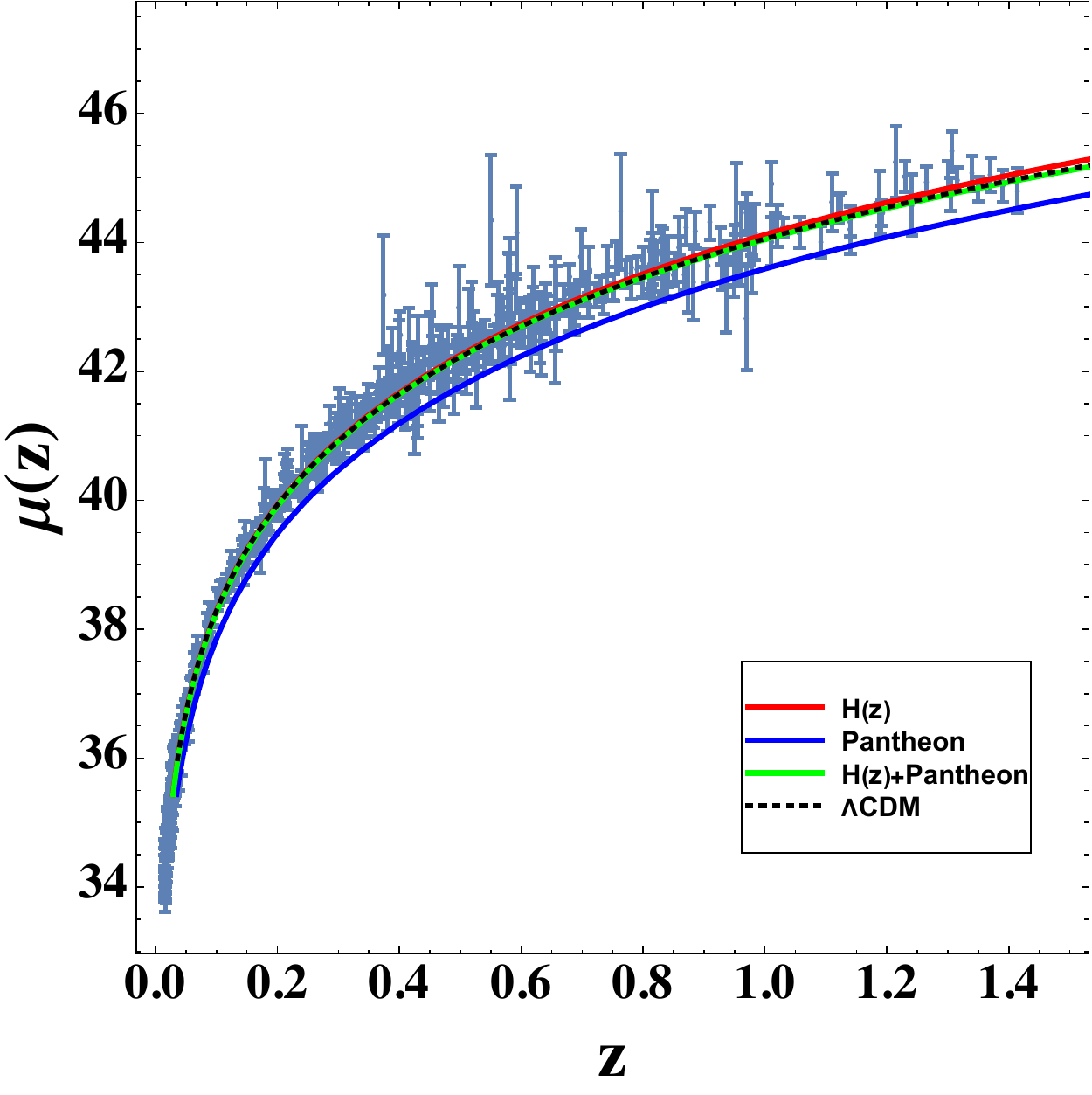}}
\caption{ The Hubble parameter $H(z)$ in terms of the redshift $z$, using various datasets, where the error bars represent the uncertainties associated with the 55 data points utilized to construct the 
Hubble datasets. The distance modulus $\mu(z)$ as a function of the redshift $z$, where the error bars represent the uncertainties associated with the Union 2.1 compilation. For completeness, in 
both graphs we have added the corresponding curves for the $\Lambda$CDM paradigm.}
 \label{H-mu}
\end{figure}

Furthermore, in  the upper graph of Figure~\ref{H-mu} we draw the Hubble parameter $H(z)$ in terms of the redshift $z$, using various datasets, while for completeness we add the corresponding curve for the $\Lambda$CDM  paradigm. The error bars depicted in the figure represent the uncertainties associated with the 55 data points utilized to construct the 
Hubble datasets. Similarly, the lower graph of Figure~\ref{H-mu} depicts the distance modulus $\mu(z)$ as a function of the redshift $z$, for the scenario at hand as well as $\Lambda$CDM cosmology, where the error bars represent  the uncertainties associated with the Union 2.1 compilation. 

\begin{figure}[!htbp]
  \includegraphics[scale=0.4]{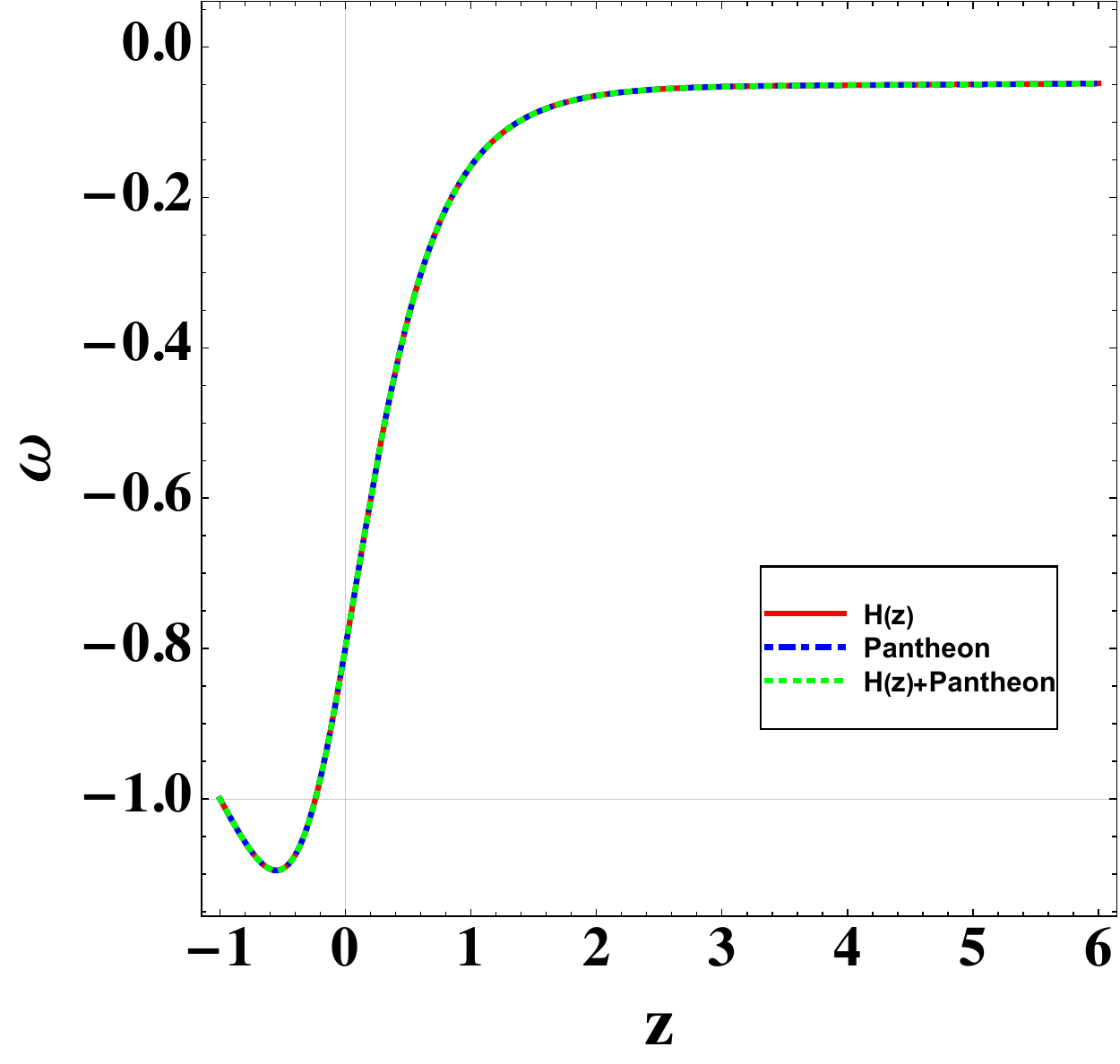}
\caption{{The evolution of the dark-energy equation-of-state parameter 
 $w_{DE}$ given by (\ref{wDEdef}), as a function of the redshift $z$,
 for the best-fit parameters  obtained from  various   datasets.   }}
\label{w-rho-p}
\end{figure}

Finally, in Figure~\ref{w-rho-p} we present the reconstructed dark-energy 
equation 
of state given by (\ref{wDEdef}). As we observe, $w_{DE}$ lies in the 
quintessence regime up to the present time, while in the future it experiences 
the phantom divide crossing. This capability was expected according to the form 
of (\ref{wDEdef}), and it was discussed in \cite{Kofinas:2014daa}.

\section{Cosmographic Analysis}
\label{section-IV}

In this section, we perform a cosmographic analysis for $f({T},{T}_{{G}})$ 
cosmology. In particular,  the dynamics of the late-time Universe is examined 
through the utilization of the Hubble parameter, deceleration parameter, and 
jerk, snap, and lerk parameters \cite{Bamba:2012cp}. Additionally, important 
information can be extracted by applying the 
 $Om(z)$ diagnostic  \cite{Sahni:2008xx}. In the following subsections, we 
investigate them in detail.

\subsection{Cosmographic Parameters}

The deceleration parameter is defined as 
\begin{equation}
 q 
= -1+(1+z)\frac{H'}{H},
\end{equation}
where   primes denote  derivatives  with respect to the redshift $ z $, and it
provides information about the acceleration rate of the 
Universe, being positive during deceleration and negative during acceleration. 
Moreover, the jerk parameter is defined as 
\cite{Visser:2004bf,Rapetti:2006fv,Liu:2023agr}
\begin{equation}
    j(z)=1-2(1+z)\frac{H'}{H}+(1+z)^2\frac{H'^2}{H^2}+(1+z)^2\frac{H''}{H}.
\end{equation}
{The} sign of the jerk parameter $j$ determines how the dynamics of the Universe 
change, with a positive value indicating the presence 
of a transition period during which the Universe modifies its expansion.
Additionally, the snap  parameter is defined as \cite{Liu:2023agr}
\begin{eqnarray}
    &&
    \!\!\!\!\!\!\!\!\!\!\!\!\!\!\!\!\!\!
    s(z)=1-3 (1\!+\!z)\frac{H'}{H}+3 (1\!+\!z)^2 
\frac{H'^2}{H^2}+(1\!+\!z)^3\frac{H'^3}{H^3}\nonumber\\
&&\! -4(1\!+\!z)^3\frac{H'H''}{H^2}
+(1\!+\!z)^2\frac {
H''}{H}-(1\!+\!z)^3\frac{H'''}{H},
\end{eqnarray}
and the lerk parameter as
\begin{eqnarray}
     && \!\!\!\!\!\!\!\!\! \! \! \! 
     l(z)=1-4 (1\!+z\!)\frac{H'}{H}+6 
(1\!+z\!)^2 
\frac{H'^2}{H^2}-4(1\!+z\!)^3\frac{H'^3}{H^3}\nonumber\\
&&+(1\!+z\!)^4\frac{H'^4}{H^4}-(1\!+z\!)^3\frac{
H'H''}{H^2}+7(1\!+z\!)^4\frac{H'H'''}{H^2}\nonumber\\
&&  
+11(1\!+z\!)^4\frac{H'^2H''}{H^3}+2(1\!+z\!)^2\frac{H''}{H}+4(1\!+z\!)^4\frac{
H''^2} {H^2 } \nonumber\\
&&
+(1\!+z\!)^3\frac{H'''}{H}+(1\!+z\!)^4\frac{H''''}{H},
\end{eqnarray}
both providing information on the higher-order derivatives of the Universe's 
acceleration, offering insights into the transitions 
between different epochs. Notably, the snap parameter 
determines the extent to which the Universe's evolution deviates from one 
predicted by the $\Lambda$CDM model.

\begin{figure}[!htbp]
 {\includegraphics[scale=0.4]{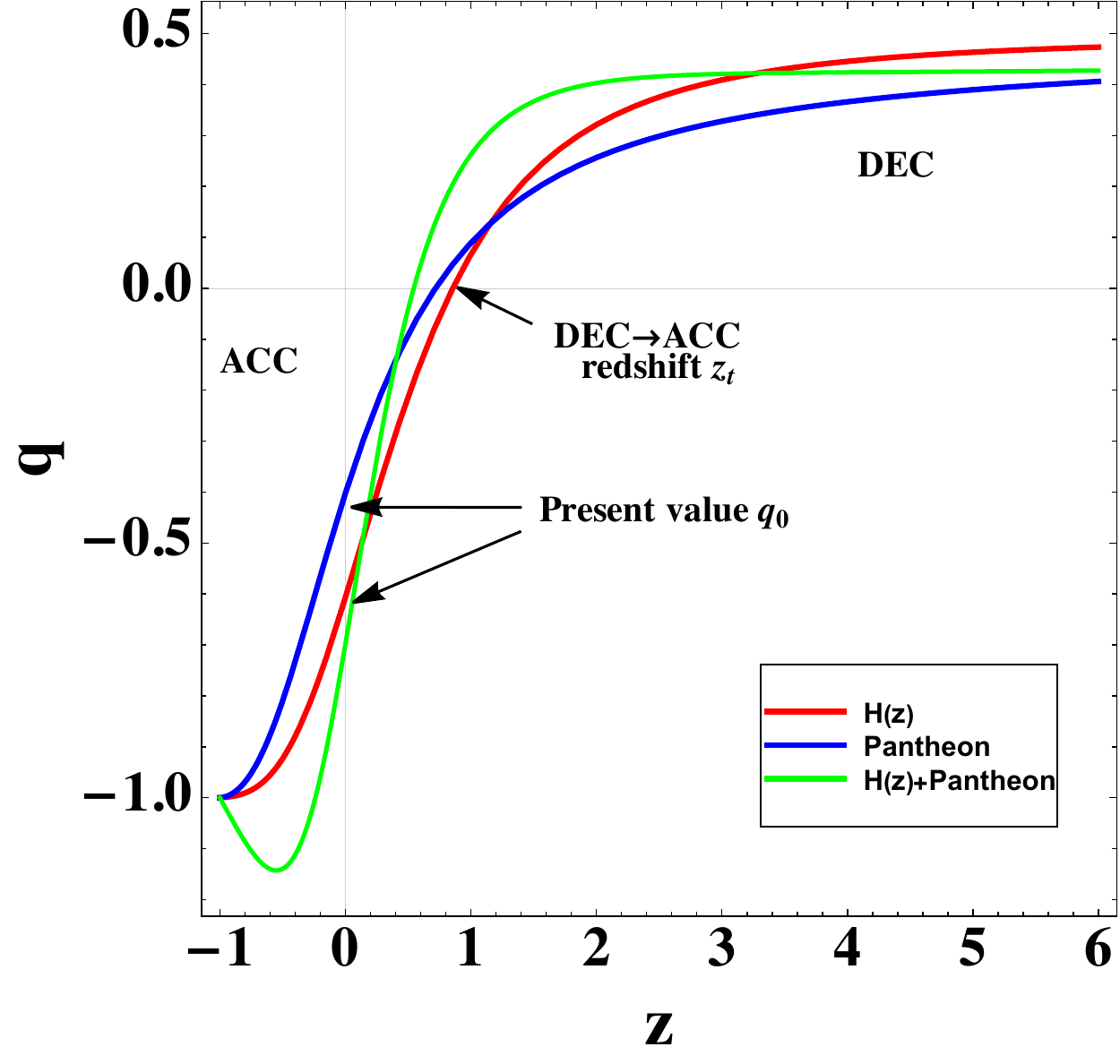}} \hfill   
 {\hspace{-1.5cm} \includegraphics[scale=0.4]{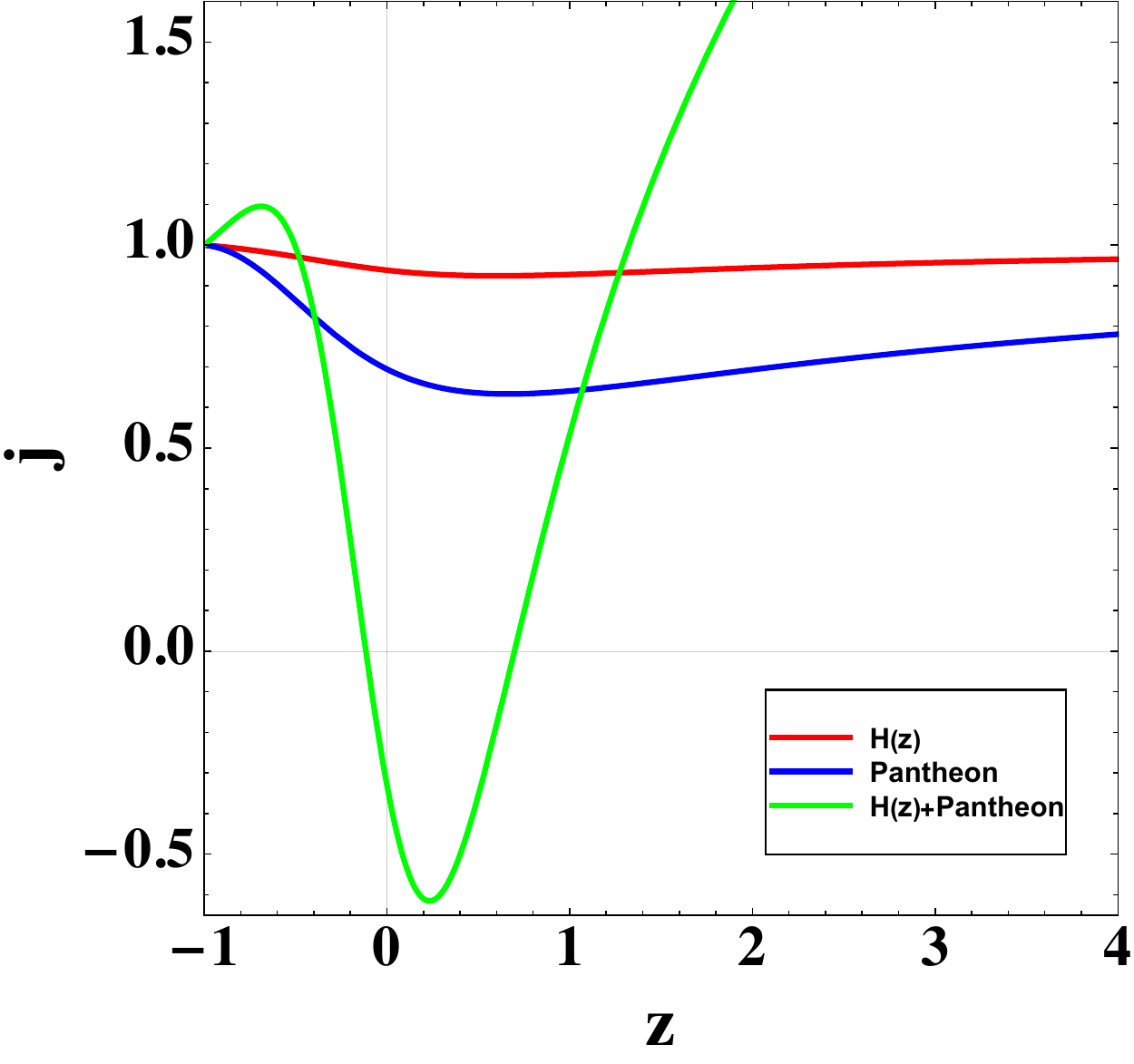}}
    \caption{ The evolution of the deceleration parameter and the evolution of the jerk parameter as a 
function of the redshift. In both graphs, we have used the best-fit parameter values obtained from various datasets.} 
  \label{dec-jerk}
\end{figure}

In the upper panel of Figure~\ref{dec-jerk}, we depict   the evolution of the 
deceleration parameter  $q$ as a function of the redshift $z$ for the model 
parameter values obtained from the observational analysis of the previous 
section. As we observe, we obtain a transition from   
deceleration to acceleration at late times in the interval $ 0.52 \leq z_{tr} 
\leq 0.89 $, as expected.
Concerning  the present-time value  $q_0$, it is found 
to be approximately $-0.60 $ and $-0.68$  using the Hubble and 
Pantheon datasets, respectively, values that are  consistent with the range of 
$q_0$ determined 
through recent observations \cite{Busca:2012bu, 
Farooq:2013hq,Capozziello:2014zda,Gruber:2013wua,Yang:2019fjt}.

The lower graph of Figure~\ref{dec-jerk}   depicts the evolution 
of the jerk parameter. As we see, the current value of $j$ is approximated to  $j_0=0.93$ and $0.69$ using the Hubble and Pantheon datasets, respectively, 
aligning with recent analyses that establish  constraints 
on the cosmographic coefficients 
\cite{Aviles2012,Aviles2014,Capozziello:2015rda}, while in the far future  
$j$   approaches 1.

\begin{figure}[!htbp]
	 { \includegraphics[scale=0.40]{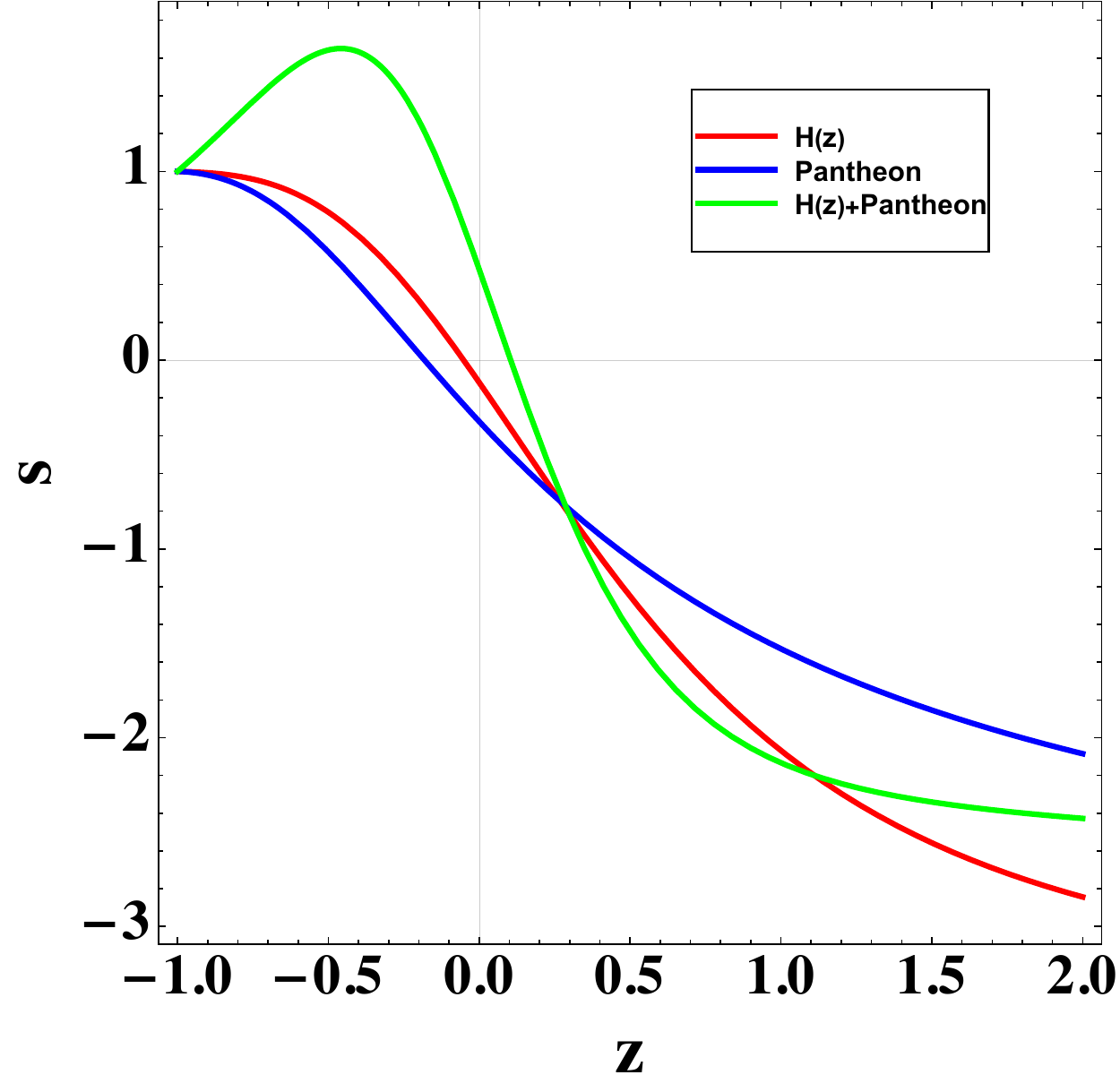}}\hfill
	 { \includegraphics[scale=0.40]{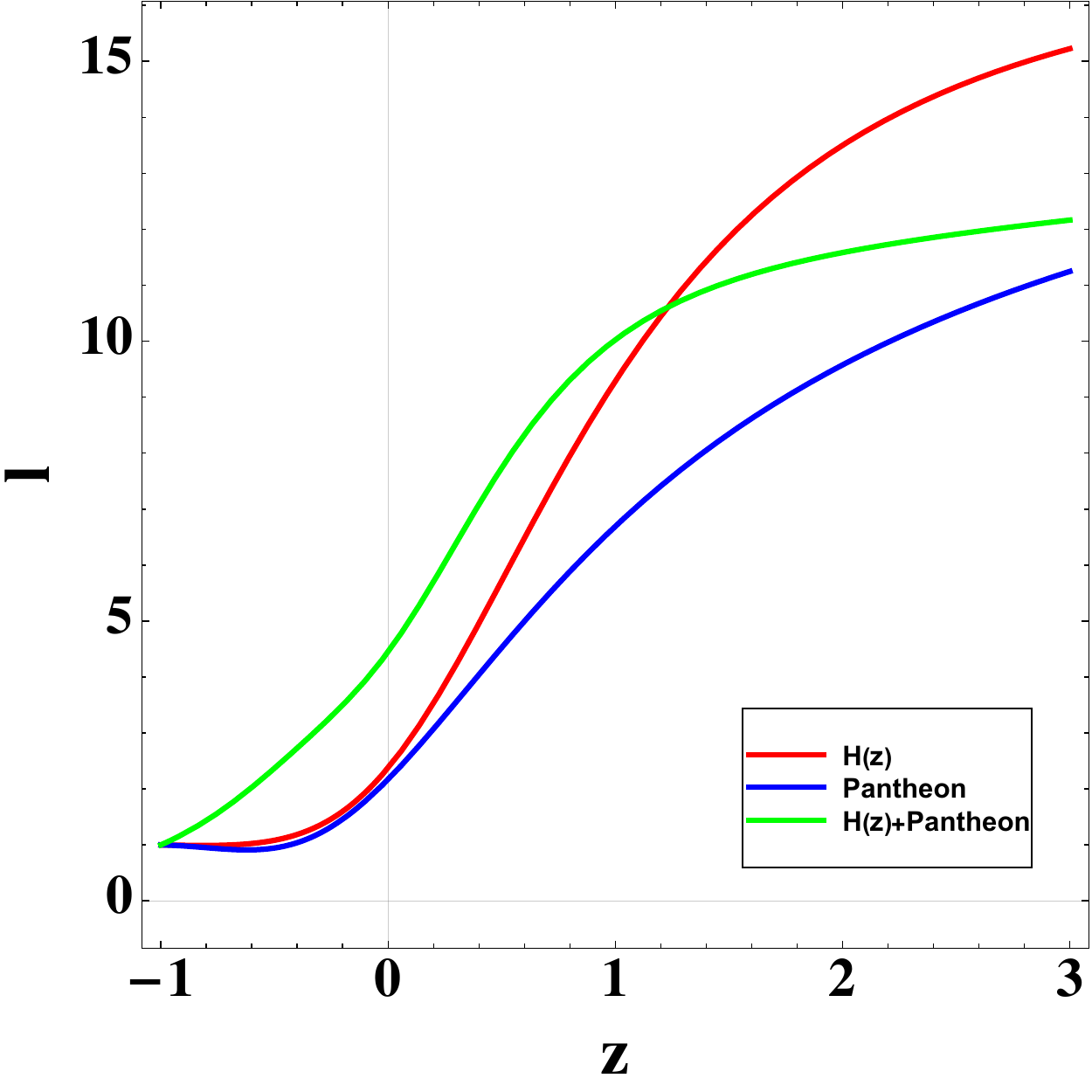}}
\caption{ The evolution of the snap parameter and the evolution of the  lerk 
parameter as a function of the redshift. In both graphs, we have used the best-fit parameter 
values obtained from various datasets. }
 \label{snap-lerk}
\end{figure}

Additionally, in the upper graph of Figure~\ref{snap-lerk} we present the 
evolution of the snap parameter $s(z)$, in which we observe a transition 
from a negative value to a positive value in the late stages.  This behavior 
aligns with the preference of the scenario for quintessence-like behavior, 
indicated by $j \leq 1$ and $s > 0$. Finally, in the lower graph of Figure 
\ref{snap-lerk} we depict the lerk parameter as a function of the redshift.
Notably, the lerk parameter rapidly decreases and stabilizes around $ \simeq$1. It is worth noting that all the geometric parameters lie within the limits 
of cosmographic coefficients determined by    late-time 
Universe  analysis \cite{Bamba:2012cp}.

\begin{figure}[!htbp]
\includegraphics[scale=0.4]{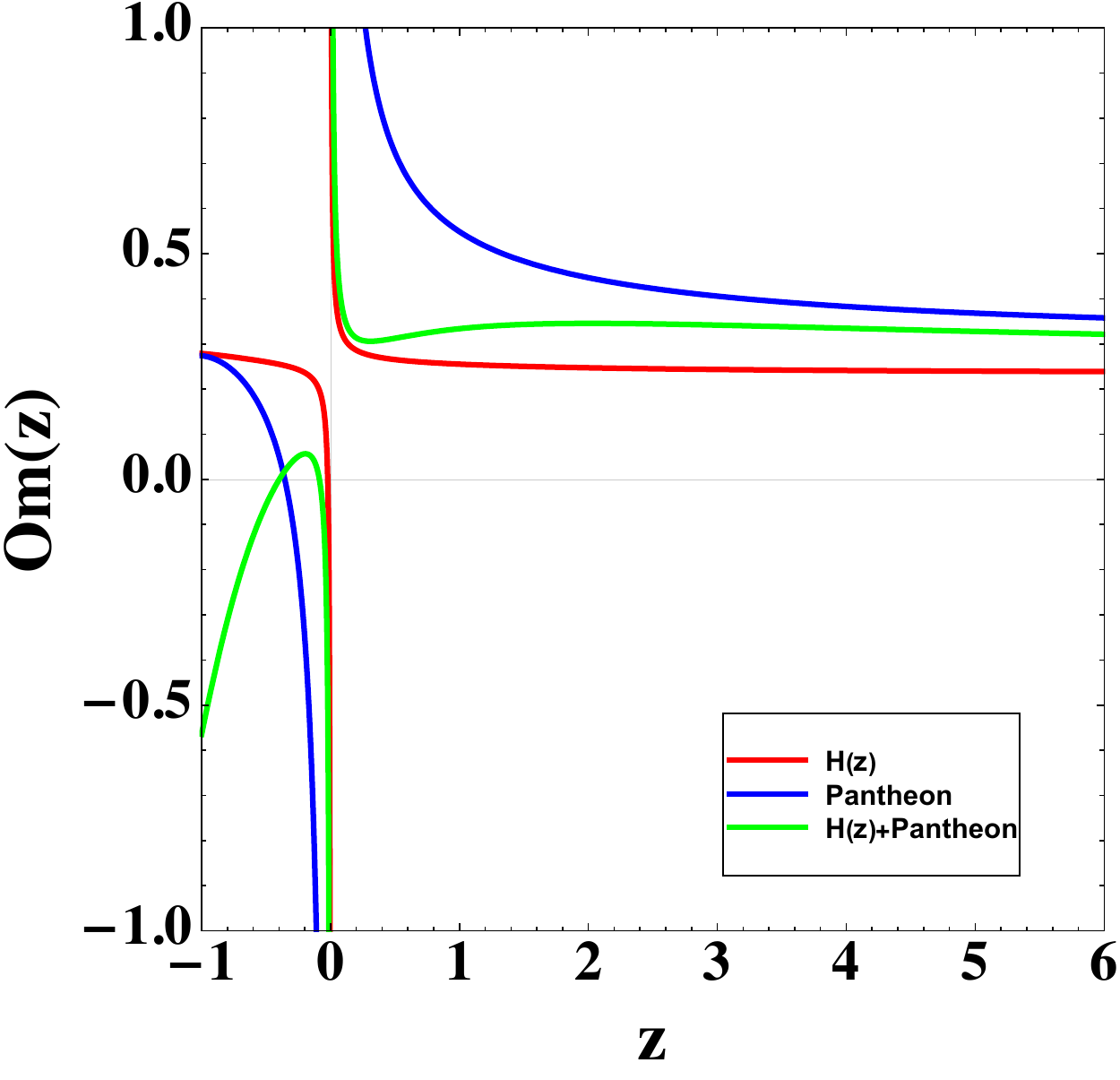}
	\caption{The evolution of the  $Om(z)$ geometrical diagnostic as a 
function of the redshift, using the best-fit parameter 
values obtained from various datasets. }
 \label{Omfig}
\end{figure}

\section{Conclusions}\label{section-V}

We performed the observational confrontation and cosmographic analysis of 
$f({T}, {T}_{{G}})$ gravity and cosmology. This higher-order torsional 
gravity is based on both the torsion scalar, as well as on the teleparallel 
equivalent of the Gauss--Bonnet combination, in its Lagrangian. Such 
a gravitational modification is different from both $f(R)$- and $F(G)$-curvature 
gravities, as well as from $f(T)$ torsional gravity, and it is known to 
have interesting cosmological applications. In particular, one obtains an 
effective dark-energy sector which depends on the extra torsion contributions.
 
Firstly, we  employed the most recent observational data obtained from the 
Hubble function  and SNeIa Pantheon datasets, in order to extract constraints 
on the free parameter space of power-law $f({T},{T}_{{G}})$ gravity. In 
particular, we applied a Markov chain Monte Carlo  sampling technique, and  we 
provided the  $1\sigma$ and $2\sigma$  iso-likelihood contours, as well as the 
best-fit values for the parameters. As we saw, the scenario at hand is in 
agreement with observations.  Additionally, we presented the 
 variations in the Hubble parameter and the distance modulu acquired  
 from the $H(z)$ dataset and the $ 1048 $ data points of Pantheon, respectively, 
where we showed that  the variations in $H(z)$ and joint datasets are closer to
$\Lambda$CDM cosmology than the Pantheon one. Finally, drawing the effective 
the dark-energy equation-of-state parameter we saw that we obtained a 
quintessence-like behavior, while in the future the Universe enters into the 
phantom regime before it   tends asymptotically to the cosmological constant 
value.
 
As a next step, we performed a detailed cosmographic analysis. From the behavior 
of the deceleration parameter, we showed that the transition from deceleration 
to acceleration occurs in the redshift range $ 0.52 \leq z_{tr} \leq 0.89 $. 
Additionally, the evolution of the jerk, snap, and lerk parameters showed the 
preference of the scenario at hand for a quintessence-like behavior. Finally, 
we applied the  $Om(z)$ diagnostic analysis, and we observed that our model 
behaves 
like a quintessence model at late times; however, in the case of the Hubble dataset 
the behavior of the model is closer to that of  $\Lambda$CDM cosmology.

The above features reveal the capabilities of $f({T},{T}_{{G}})$ gravity and 
cosmology.   The physical aspects of these 
models  have been discussed in 
detail in the literature; however, the novel result of the present analysis 
is the complete observational confrontation of these theories with the 
latest datasets.
Nevertheless, many tests need to be performed before the 
scenario can be considered a good candidate for the description of nature. 
A necessary investigation is the detailed examination of the perturbations, and  
the study of the matter overdensity growth, since this will allow us to compare 
the scenario with data from large-scale structure, such as $f\sigma_8$, 
and 
other perturbation-related probes.  Furthermore, one should investigate the 
constraints on the theory from solar system experiments, since they are known 
to be very restrictive. These studies lie beyond the scope of the present work 
and are left for future projects.

\begin{acknowledgments}
 
ENS acknowledges the contribution of the LISA CosWG and of COST Actions  CA18108  ``Quantum Gravity Phenomenology in the multi-messenger approach''  and  CA21136 ``Addressing observational tensions in cosmology with 
systematics and fundamental physics (CosmoVerse)''. The authors express their thanks to the reviewers for their valuable comments and suggestions which improved the research quality and presentation significantly.
 
\end{acknowledgments}

\end{document}